\documentclass[10pt,a4paper]{article}
\RequirePackage{latexsym,amsmath,amssymb}
\RequirePackage[dvipsnames,usenames]{color}

\usepackage{fullpage}
\usepackage{cite}

\usepackage[english]{babel}
\usepackage[T1]{fontenc}
\usepackage[final]{showkeys} % [draft/final]
\usepackage[hidelinks]{hyperref}
\hypersetup{pdfauthor={M. Cadoni, M. Oi, A. P. Sanna},
            pdftitle={}
            }
\usepackage[]{cleveref}
\usepackage{amsmath}
\usepackage{ mathrsfs }
\usepackage{subcaption}
\usepackage{cancel}
\usepackage{url}

\usepackage{tikz}
\usetikzlibrary{decorations.pathmorphing}

\Crefname{equation}{Eq.}{Eqs.}
\Crefname{figure}{Fig.}{Figs.}
\crefformat{plural}{#2eqs.~(#1)#3}
\crefname{section}{Sect.}{Sects.}

\usepackage{amsfonts}
\usepackage{graphicx,epsfig}
\def\be#1\ee{\begin{align}#1\end{align}}

\renewcommand{\ge}{\geqslant}

\def\a{\alpha}

\def\lb{\label}

\def\beq{\begin{equation}}
\def\eeq{\end{equation}}
\newcommand{\rH}{r_{\text{H}}}
\def\a{\alpha}

\title{\bf Quasi-normal modes and microscopic description of 2D black holes }
\author{M.~Cadoni${}^{ab}$\thanks{E-mail: mariano.cadoni@ca.infn.it}, \ M.~Oi${}^{ab}$\thanks{E-mail: mauro.oi@ca.infn.it}, \ 
A. P. ~Sanna${}^{ab}$\thanks{E-mail: asanna@dsf.unica.it} \ 
\\
${}^a$\emph{Dipartimento di Fisica, Universit\`a di Cagliari}
\\
{\em Cittadella Universitaria, 09042 Monserrato, Italy}
\\
\\
${}^b$\emph{I.N.F.N, Sezione di Cagliari}
\\
{\em  Cittadella Universitaria, 09042 Monserrato, Italy}
\\
\\}
\begin{document}
\maketitle
\begin{abstract}
We investigate the possibility of using quasi-normal modes (QNMs) to probe the microscopic structure of  two-dimensional (2D) anti-de Sitter (AdS$_2$) dilatonic black holes. We first extend previous results on the QNMs spectrum, found for external massless scalar perturbations, to the case of massive scalar perturbations. We find that the quasi-normal frequencies are purely imaginary and scale linearly with the overtone number. Motivated by this and extending previous results regarding Schwarzschild black holes, we propose a microscopic  description of the 2D black hole in terms of a coherent  state of $N$ massless particles quantized on a circle, with occupation numbers sharply peaked on the characteristic QNMs frequency $\hat \omega$. We further model the black hole as a statistical ensemble of $N$ decoupled quantum oscillators of frequency $\hat\omega$. This allows us to recover the Bekenstein-Hawking (BH) entropy $S$ of the hole as the leading contribution to the Gibbs entropy for the set of oscillators, in the high-temperature regime, and to  show that $S=N$. Additionally, we find sub-leading logarithmic corrections to the  BH entropy. We further corroborate this microscopic description by outlining a holographic correspondence between QNMs in the AdS$_2$ bulk and the de Alfaro-Fubini-Furlan conformally invariant quantum mechanics. Our results strongly suggest that modelling a black hole as a coherent state of particles and as a statistical ensemble  of decoupled harmonic oscillators is always a good approximation in the large black-hole mass, large overtone number limit.  
\end{abstract}
%
%\end{center}

%\end{titlepage}

\newpage
\tableofcontents

\section{Introduction}
\lb{secint}
In recent times, we have gained growing theoretical evidence that some peculiar and puzzling features of the gravitational  interaction could be addressed by assuming the underlying microscopic quantum theory of gravity to have a {\it multi-scale}  behavior, with the generation of different infrared (IR) length-scales,  both at the black hole horizon  and at galactic  and cosmological level \cite{Verlinde:2016toy,Cadoni:2017evg, Tuveri:2019zor, Dvali:2020wqi, Bose:2021ytn}. The holographic principle, the information paradox for black holes, the deviation from Newtonian  dynamics at galactic scales and the origin of dark energy  stand out among these puzzles.  

Evidence along this direction came out few years ago from the proposal to describe   a black hole with mass $M$ and  radius  $R_\text{S}=2GM$  as a condensate of a large number $N_\text{g}\sim \ell_\text{P}^2 M^2$ (where $\ell_\text{P}=\sqrt G$ is the Planck length\footnote{We adopt natural units, $c=\hbar =1$.}) of soft gravitons with typical energy $\epsilon\sim 1/R_\text{S}$ \cite{Dvali:2011aa, Dvali:2012en, Dvali:2013eja, Cadoni:2020mgb, Dvali:2012rt}. This is tightly connected to the {\it classicalization} idea,  according to which gravity achieves  UV-completion by using classical objects of  size  $R_\text{S}\gg  \ell_\text{P}$, but composed by a huge number of weakly coupled soft quanta of momenta $p\sim 1/R_\text{S}$  \cite{Dvali:2011aa, Dvali:2010jz, Dvali:2011th}.
In this way, a strong 't~Hooft-like coupling $\lambda= \alpha N_\text{g}$  (with $\alpha$ the running coupling) is traded for a large number of weak interacting particles, generating an IR length-scale $\ell\sim \ell_\text{P} \sqrt{N_\text{g}}$ \cite{Dvali:2020wqi, Bose:2021ytn, Dvali:2011aa}.

The same idea can be applied   at  cosmological and  galactic scales. Indeed, in the former case, the de~Sitter universe can be seen as a bound state of a huge number of soft particles with typical energy  $\epsilon\sim 1/L$, with $L$ the size of the cosmological horizon. In the latter case, long-range quantum gravity  effects could generate an IR length-scale of galactic size $r_0= \ell_\text{P} \sqrt{N_\text{G}}$,  at which the dynamics deviates from the Newtonian behavior \cite{Verlinde:2016toy, Cadoni:2017evg, Tuveri:2019zor}. Here, $N_\text{G}\sim \ell_\text{P}^2 M^2_\text{G}$ is the number of particles associated with the mass $M_\text{G}$ of the galaxy.    

A quite similar  black hole description emerged also in the string-theory context, and led to the fuzzball proposal \cite{Lunin:2001jy, Lunin:2002qf, Mathur:2005zp,Ikeda:2021uvc}. Specifically, a black hole is considered as a bound state of a large number  of extended objects (fuzzballs) of size $R\sim R_\text{S}+\ell_\text{P}$.  These configurations receive contributions also from virtual fluctuations of the quantum gravity vacuum, dubbed as VECROs \cite{Mathur:2019dhf, Mathur:2020ely}.

Altogether, these results may be seen as a strong indication of the existence of an internal, non-trivial, microscopic black hole structure, which should represent  the starting point  for explaining the black hole  macroscopic behavior. Unfortunately, owing to the classical no-hair theorems, it is almost impossible for a distant observer to have direct access to this internal microscopic information. The only direct window an asymptotic observer has on the microscopic black-hole structure is the way the hole responds to external perturbations. This response is codified in the quasi-normal modes (QNMs) spectrum \cite{Berti:2009kk, Kokkotas:1999bd, Konoplya:2011qq}.  QNMs are the characteristic oscillations produced by a perturbed black hole and they decay exponentially in time. Their spectrum   
can be experimentally detected, by observing the gravitational  wave signal originated in the ringdown phase of two compact objects  merging to form a black hole. 
 
QNMs correspond to complex eigenfunctions of the linearized  system. They are  characterized by complex frequencies, where the imaginary part describes the damping of the mode caused by purely ingoing boundary conditions at the horizon. Moreover, Dirichlet boundary conditions at infinity   imply a discrete spectrum for the  frequencies  $\omega = \omega_\text{R}+ \text{i}\omega_\text{I}$, with the imaginary part depending on an integer $n$, the overtone number.

Despite QNMs being classical, there has been strong indications that they could contain information about the quantum properties of black holes \cite{Maggiore:2007nq}, their horizons and their internal microscopic structure. In particular, this is true  in the high-damped (large-$n$) regime, when QNMs are expected to probe the black hole at very short distances.
 
The high-damped QNMs spectrum, together with an analogue of Bohr's correspondence principle, has been used by Hod to explain and fix the area spectrum of the event horizon (see, e.g. Refs. \cite{Hod:1998vk, Dreyer:2002vy, Kunstatter:2002pj, Corda:2015rna}),  whose quantization was first suggested by Bekenstein in Refs. \cite{Bekenstein:1974jk, Bekenstein:1995ju}. The QNMs spectrum has been also used  in  the  Loop Quantum Gravity framework   to fix  the Barbero-Immirzi parameter \cite{Dreyer:2002vy}, which is essential to correctly account for the Bekenstein-Hawking (BH) entropy \cite{Immirzi:1996dr, Agullo:2008yv}.

The  next, crucial, step along this  direction was done  by Maggiore \cite{Maggiore:2007nq}, who showed that the QNMs spectrum for the Schwarzschild  black hole, in the high-damping limit, can be interpreted as a damped harmonic oscillator, with proper frequency $\omega = \sqrt{\omega_\text{R}^2 + \omega_\text{I}^2}$. Moreover, he argued that this result corresponds to the dispersion relation of a massive particle quantized on a circle with length given by the inverse of the black-hole Hawking temperature.

In a previous paper \cite{Cadoni:2021jer}, building on Maggiore's result, we have proposed a microscopic description of the Schwarzschild black hole in terms of  a canonical ensemble of $N$ decoupled harmonic oscillators. Using this model, we have reproduced the BH entropy as the leading contribution to the Gibbs entropy, in the large-mass black hole limit. We have also derived subleading, logarithmic corrections to the BH result, in agreement with several results in the literature. We additionally found that the number of oscillators scales holographically with the area of the event horizon.

The natural question that now arises is how general the results of Ref. \cite{Cadoni:2021jer} are. Do they represent a peculiarity of the Schwarzschild black hole or are they, instead, a feature common to all black holes? Since the imaginary part of $\omega$ always scales linearly with $n$ in the high-damping regime, we expect the main results of Ref. \cite{Cadoni:2021jer} to generally hold\footnote{A general description of a black hole as a set of quantum harmonic oscillators has been also proposed  in Ref. \cite{Casadio:2013hja}, by modelling the black hole as a spherical cavity.}.

In this paper, we tackle this issue by considering a two-dimensional (2D) dilaton gravity model, namely Jackiw-Teitelboim  (JT)  gravity \cite{Jackiw:1984je, Teitelboim:1983ux, Grumiller:2002nm}.  Despite being particularly simple, this model allows for AdS$_2$ black holes, i.e 2D  solutions with anti de Sitter  asymptotic behavior. Because of its simplicity, the model has been used in several contexts \cite{Grumiller:2002nm}, like for instance the Hawking evaporation process \cite{Cadoni:1994uf, Cadoni:1995dd, Kim:1999un, Engelsoy:2016xyb, Mertens:2019bvy} and the related information puzzle \cite{Almheiri:2019psf, Almheiri:2019hni, Almheiri:2019yqk, Penington:2019kki, Almheiri:2019qdq, Goto:2020wnk, Verheijden:2021yrb, Cadoni:2021ypx}, AdS/CFT correspondence and computation of entanglement entropy; it is also closely related to the 
Sachdev-Ye-Kitaev (SYK) model \cite{Kitaev:2017awl}.

We first extend previous results for the QNMs spectrum of the JT black hole \cite{Bhattacharjee:2020nul} to include the case of massive scalar perturbations. We show that, for generic (massless or massive) scalar perturbations, the quasi-normal frequencies are purely imaginary and scale linearly with the overtone number.

We then show that the QNMs spectrum agrees with the dispersion relation of a massless particle, quantized on a circle of length given by half of the inverse Hawking temperature of the hole. We use the same approach of Ref. \cite{Cadoni:2021jer} to model the JT black hole first  as coherent state with occupation numbers sharply peaked on the characteristic QNMs frequency $\hat\omega$, and then as a statistical ensemble of $N$ decoupled quantum harmonic oscillators of frequency $\hat{\omega}$. This latter description enables us to derive the BH entropy of the hole as the leading contribution to the Gibbs entropy, in the high-temperature regime. Sub-leading corrections are also computed and shown to behave logarithmically. Further, we find that $N$ equals the BH entropy. 

Furthermore, apart from its simplicity, the JT black hole is interesting because it allows for a dual description in terms of a conformal field theory (CFT).  Motivated by this, we investigate the relationship between our description in terms of a set of harmonic oscillators and the dual CFT. We find a natural holographic correspondence between QNMs in the AdS$_2$ bulk and  de Alfaro-Fubini-Furlan (DFF) conformally invariant quantum mechanics, which gives a conformally-invariant generalization of the usual quantum harmonic oscillator.

The present work is organised as follows.
In \cref{sect2} we briefly review the main features of black-hole solutions in JT gravity.
In \cref{Sec:QNMSAnalyticalSolutions} we review the calculations of the  QNMs spectrum for external massless scalars and extend these to massive scalar perturbations.  
The microscopic descriptions of the JT black hole in terms of quantum particles and a canonical ensemble of decoupled harmonic oscillators are presented in \cref{Sec:cd,Sec:BHQuantumOscillator}, respectively.
In \cref{Sec:Mapto1DQMmodel} we discuss the holographic correspondence between QNMs in the AdS$_2$ bulk and  DFF conformal quantum mechanics.
Finally in \cref{sect7} we present our conclusions.   

\section{2D  AdS black holes}
\lb{sect2}
In this paper we consider AdS$_2$ black-hole solutions of Jackiw-Teitelboim (JT) gravity, a well-known gravity theory in two space-time dimensions, described by the action
\begin{equation}
\label{JTnoanomaly}
\mathcal{S}_{\text{JT}}=\frac{1}{2\pi} \int d^2 x \sqrt{g} \ \phi \left(R+2\Lambda^2\right),
\end{equation}
where $\phi$ is a scalar field (the dilaton), $R$ the 2D Ricci scalar, $2\Lambda^2$ the cosmological constant. In the Schwarzschild gauge, where the line element can be written as $ds^2 =-f(r) dt^2 + dr^2/f(r)$, the vacuum solutions for the metric and the dilaton read 
\begin{equation}
\label{staticblackholesolSchw}
ds^2 =-\left(\frac{r^2}{L^2}-a^2 \right)dt^2+\left(\frac{r^2}{L^2}-a^2 \right)^{-1}dr^2,\qquad \phi(r) =  \phi_0\frac{r}{L},
\end{equation}
where $L=1/\Lambda$ is the AdS length, $\phi_0$ and $a^2$ are integration constants, parametrizing the solutions. Since $\phi_0$ is not relevant for our discussion, we will set it to $1$ for simplicity in the following. The Arnowitt-Deser-Misner (ADM) mass of the solution, therefore, takes the form \cite{Cadoni:1994uf, Cadoni:1993rn} 
\begin{equation}
    \label{ADMmass}
M=\frac{a^2}{2 L}.
\end{equation}
In this paper, we will only consider solutions with $a^2 \ge 0$, which have positive ADM mass and a Killing horizon at $r = \rH = a L$, and describe asymptotically AdS black holes, with an event horizon at $r=\rH$ and a singularity at $r=0$ \cite{Cadoni:1994uf}.

Using the standard procedure, one easily finds the temperature $T_\text{H}$ and the entropy $S$ associated with the black hole \cite{Cadoni:1993rn} 
\begin{subequations}
\begin{align}
&T_\text{H} =  \frac{r_\text{H}}{2\pi  L^2}=\frac{1}{2\pi} \sqrt{\frac{2M}{L}}, \label{Temperature}\\
&S = 4\pi  \sqrt{\frac{M  L}{2}} = 2\pi \frac{r_\text{H}}{L}. \label{Entropy}
\end{align}
\end{subequations}
To conclude this section, let us briefly discuss the causal structure of AdS$_2$. The maximal extension of our black hole geometry,  the full AdS$_2$ space-time, is regular and has two disconnected parts. In the Penrose diagram, the black-hole geometry represents only one of the two wedges building the  maximally extended AdS$_2$ up \cite{Cadoni:1994uf}. We will use this causal structure in \cref{Sec:BHQuantumOscillator}, when dealing with the microscopic derivation of the black hole entropy.
  
\section{Scalar perturbations and quasi-normal modes}
\label{Sec:QNMSAnalyticalSolutions}

Quasi-normal modes (QNMs) represent the characteristic oscillations of a black hole reacting to external perturbations. In the linear regime, one finds that QNMs are characterized by a discrete spectrum of complex frequencies, whose imaginary part describes the damping of the mode in time, once appropriate boundary conditions at infinity and at the horizon are imposed (see, e.g. Refs. \cite{Berti:2009kk, Kokkotas:1999bd, Konoplya:2011qq}). 

The 2D case is peculiar since pure Einstein gravity is topological in two space-time dimensions. In JT gravity, the presence of a further degree of freedom (DOF), i.e. the dilaton, allows for the existence of global modes, but not for vectorial nor tensorial propagating ones. Thus, in this theory, QNMs are linked only to external scalar perturbations of the black hole space-time \eqref{staticblackholesolSchw}. The QNMs spectrum for massless scalar perturbations in 2D-gravity has been already investigated in the literature, by imposing either usual boundary conditions or monodromy conditions around the black-hole horizon \cite{ Bhattacharjee:2020nul, Kettner:2004aw, LopezOrtega:2011np, Cordero:2012je, Hernandez-Velazquez:2021zoh}. In this section, we will briefly review the relevant results and extend them to the case of massive scalar perturbation.       

Consider a perturbing, external, massless scalar field $\Phi$ in the black hole space-time \eqref{staticblackholesolSchw}. This perturbation satisfies a generalized Klein-Gordon equation
\begin{equation}\label{KG}
    \frac{1}{\sqrt{-g}\,h(\phi)}\partial_\mu\left(\sqrt{-g}\,h(\phi)g^{\mu\nu}\partial_\nu\Phi\right)=0,
\end{equation}
where  $h(\phi)$ is a generic arbitrary coupling between the dilaton and the perturbing field $\Phi$. Since the  background dilaton solution is a function of the radial coordinate (see \cref{staticblackholesolSchw}), we have $h(\phi)\equiv h(r)$. The presence of the coupling $h(\phi)$ is  justified in a general setting. Moreover, it is fully natural if one considers the JT black hole as a dimensional reduction of higher-dimensional models-- and in particular of the 3D BTZ black hole \cite{Achucarro:1993fd}. In this case, $h(\phi)$ encodes the information on the higher-dimensional theory. 

Decomposing $\Phi(t,r)$ into Fourier modes, we can redefine the perturbing field as 
\begin{equation}\label{ansatzpertfield}
    \Phi(t, r)=e^{\text{i}\omega t}\frac{R(r)}{\sqrt{h(r)}}.
\end{equation}
\Cref{KG} can then be written in terms of the radial function $R(r)$ only, which satisfies
\begin{equation}\label{radialperteq}
    \frac{d^2R}{dr_\ast^2}+\left[\omega^2-V(r)\right]R=0,
\end{equation}
where we introduced the tortoise coordinate $r_\ast$, defined by $dr_\ast/dr=L^2/(r^2-\rH^2)$ or, equivalently, by $r =-\rH \coth\left( \rH r_*/L^2 \right)$. This transformation  maps the horizon $r=\rH$ into $r_\ast \to -\infty$ and spatial infinity $r\to \infty$ into $r_\ast =0$. Moreover, the potential $V(r)$ is given by
\begin{equation}\label{pertpotential}
    V(r)=\frac{f(r)}{2h(r)}\left[f(r)\frac{d^2h}{dr^2}+\frac{df}{dr}\frac{dh}{dr}-\frac{f(r)}{2h(r)}\left(\frac{dh}{dr}\right)^2\right].
\end{equation}

We adopt usual boundary conditions for QNMs requiring Dirichlet conditions at infinity, i.e. the radial function has to behave as $R(r)\sim 0$ at $r\to\infty$,  and  we must have purely ingoing modes at the horizon. \Cref{radialperteq}, together with boundary conditions, represents  an eigenvalue problem in the frequency $\omega$, which can be solved once the function $h(\phi)$ is chosen. 

As expected in view of the non-existence of vectorial and tensorial propagating modes in pure JT gravity, a non-trivial coupling function between the scalar field $\Phi$ and the dilaton is required to have non-trivial propagating solutions. In fact, when $h(r)=\text{constant}$, the potential $V(r)$ vanishes and \cref{radialperteq} is solved by a freely propagating perturbation, which is not compatible with the QNMs boundary conditions.

The  most natural choice for the coupling function $h$ is a power law in $\phi$  (hence in $r$): $h(r)=(r/L)^\alpha$. This is motivated not only by simplicity arguments, but also by regarding JT gravity as the dimensional reduction of a $d+2$ dimensional theory. Indeed, in the latter case, we have $h(r)=\phi(r)=r/L$ \cite{Kettner:2004aw, Kunstatter:1997my}. Adopting the  power-law form for $h$, the potential \eqref{pertpotential} becomes 
\begin{equation}\label{eq:V_alpha}
    V(r) = \frac{1}{4} \alpha  (\alpha +2)\frac{ r^2}{L^4}+\frac{(\alpha -2) \alpha   \rH^4}{4 L^4r^2}- \alpha ^2 \frac{ r_\text{H}^2}{2L^4} ,
\end{equation}
which, in terms of the tortoise coordinate, reads 
\begin{equation}\label{Vrast}
    V(r_\ast)=\frac{\rH^2\a}{4 L^4 \sinh^2 \left(\rH r_\ast/L^2\right) \cosh^2 \left(\rH r_\ast/L^2 \right)}\left[\a + 2 \cosh\left(2 \rH r_\ast /L^2\right) \right].
\end{equation}

As already mentioned at the beginning of the section, the QNMs spectrum for $\omega$ can be obtained either by solving \cref{radialperteq} with usual boundary conditions at infinity and at the horizon \cite{Cardoso:2001hn}, or by imposing a monodromy condition around the black hole horizon (see, e.g. Ref. \cite{Kettner:2004aw}). In the following, we will briefly review the computations of Ref. \cite{Cardoso:2001hn} and then we will apply them to derive the frequency spectrum for scalar perturbations.

The Klein-Gordon equation \eqref{radialperteq}, together with the potential \eqref{Vrast}, can be recast in a more suitable form by introducing a new radial coordinate $x=1/\cosh( \rH r_\ast/L^2)$. This coordinate change maps the horizon and spatial infinity respectively to $x=0$ and $x=1$. Moreover, it is useful to factorize the asymptotic behavior of $R(x)$ in these two limits. Since boundary conditions require $R(x)$ to be purely ingoing at the horizon and to vanish at spatial infinity, one can write  $R(x)=(x-1)^{(\alpha+2)/4}x^{-\text{i}  L^2 \omega/2\rH}F(x)$, so that \cref{radialperteq}, once written in terms of $F(x)$, takes the form%
\begin{equation}
    x(1-x) \frac{\text{d}^2F}{\text{d}x^2}%
    +\left[1-\frac{\text{i} \tilde \omega }{\rH}-\frac{1}{2} \left(\alpha -\frac{2 \text{i} \tilde \omega }{\rH}+5\right)x\right]\frac{\text{d}F}{\text{d}x}%
    -\frac{\left(2 \rH-\text{i} \tilde \omega \right) \left[(\alpha +1)  \rH-\text{i} \tilde \omega \right]}{4\rH^2}F(x)=0,
\end{equation}
where  $\tilde \omega= L^2 \omega$. The solution of this equation is a combination of hypergeometric functions. By selecting purely ingoing modes at the horizon, we are left with
\begin{align}\label{eq:solution}
    F(x) &= \mathcal{C}\, {}_2F_1\left(1-\frac{\text{i}\tilde \omega}{2\rH},\frac{1+\alpha}{2}-\frac{\text{i}\tilde \omega}{2\rH},1-\frac{\text{i}\tilde \omega}{\rH},x\right)=\nonumber\\%
    &=\mathcal{C} (1-x)^{-\frac{\alpha+1}{2}} \, {}_2F_1\left(-\frac{\text{i} \omega }{2 \rH},\frac{1-\alpha }{2}-\frac{\text{i} \omega }{2 \rH},\frac{1 + \alpha}{2}-\frac{\text{i} \omega }{2 \rH}, x\right),
\end{align}
where $\mathcal{C}$ is a constant. \Cref{eq:solution} still does not automatically satisfy boundary conditions at infinity. Requiring
\begin{equation}\label{eq:quantization_od_QNMs}
    {}_2F_1\left(-\frac{\text{i}\,\tilde \omega}{2\rH},\frac{1-\alpha}{2}-\frac{\text{i}\,\tilde \omega}{2\rH},\frac{1+\alpha}{2}-\frac{\text{i}\,\tilde \omega}{2\rH},1\right)=%
    \frac{\Gamma \left(\frac{1+\alpha }{2}\right) \Gamma \left(1-\frac{\text{i} \tilde \omega }{\rH}\right)}{\Gamma \left(1-\frac{\text{i} \tilde \omega }{2 \rH}\right) \Gamma \left(\frac{1+\alpha }{2}-\frac{\text{i} \tilde \omega }{2 \rH}\right)}=0,
\end{equation}
and by solving \cref{eq:quantization_od_QNMs}, we get the desired behavior. \Cref{eq:quantization_od_QNMs} is satisfied only when $\alpha$ is a non-even real number and $\omega$ is 
\begin{equation}\label{freqmassless}
    \omega\equiv \omega_n = -\frac{2\text{i}\rH}{L^2}\left(n+\frac{\alpha+1}{2}\right),
\end{equation}
where $n$ is the overtone number, an integer labelling the mode.

Notice that the quasi-normal frequencies are purely imaginary and the resulting modes purely damped. This fact derives from the asymptotic $\text{AdS}_2$ behavior of the JT black hole, which is reflected in the form of the potential and in the related boundary conditions. 

\subsection{Quasi-normal modes for massive scalar perturbations}
\label{Sec:AnalyticMeth}

Previously, we considered the evolution of external massless scalar perturbations in the fixed gravitational background \eqref{staticblackholesolSchw}. Let us now consider external  perturbations due to a scalar field $\Phi$  of mass $m$ and a coupling function with the dilaton $h(\phi)$. The equation for the perturbation is now 
\begin{equation}\label{KGdilaton}
    \left[\frac{1}{\sqrt{-g}\,h(\phi)}\partial_\mu\left(\sqrt{-g}\,h(\phi)g^{\mu\nu}\partial_\nu\right)- m^2\right]\Phi=0.
\end{equation}
We will consider the same coupling function $h(\phi)=(r/L)^\alpha$  used for the massless case. Our results will therefore generalize those in Ref. \cite{Cordero:2012je}, derived considering a trivial coupling $h(\phi)=\text{constant}$. 
This choice,  together with the ansatz \eqref{ansatzpertfield}, leads to the equation
\begin{equation}
    \frac{d^2R}{dr_\ast^2}+\left[\omega^2 - V(r) - m^2(r^2-\rH^2)\right]R=0,
\end{equation}
where $V(r)$ has been defined in \cref{eq:V_alpha}.
Following the same procedure described in the previous subsection when dealing with the massless case, we get the spectrum of QNMs for massive scalar perturbations

\begin{equation}\label{freqdilaton}
    \omega_n = -\frac{2\,\text{i}\rH}{L^2}\left(n+\frac{1+\alpha}{4}+\frac{1}{4}\sqrt{(1+\alpha)^2+4L^2m^2}\right).
\end{equation}%

%.
The results of this section show that, for both massless and massive scalar perturbations, the QNMs spectrum of JT  black holes takes the general form $\omega_n = -\frac{2\text{i}\,\rH}{L^2}\left(n+\gamma\right)$, where $\gamma$ is some real number  of order one. The frequencies are purely imaginary and grow linearly with the overtone number $n$. This latter behavior is reminiscent of the energy spectrum of a one-dimensional quantum harmonic oscillator \cite{Maggiore:2007nq}. 

In the high-damping regime, we can neglect $\gamma$. The relevant information here is the linear scaling of $\omega_n$ with  $n$. We can now exploit Maggiore's argument, originally used for the Schwarzschild black hole \cite{Maggiore:2007nq}, to suggest a correspondence between the QNM frequencies $\omega_\text{R} + \text{i}\omega_\text{I}$ and the proper frequency $\omega$ of an harmonic oscillator via the relation $\omega = \sqrt{\omega_\text{R}^2 + \omega_\text{I}^2}$. In the present case, this gives
\begin{equation}\label{freqQNMs}
    \omega_n \equiv |\omega_\text{I}| =\hat{\omega}\left(n+\gamma \right),\,\, \hat{\omega}=\frac{2\rH}{L^2},
\end{equation}
where the  characteristic frequency $\hat{\omega}$ if defined in terms of the fundamental frequency of the oscillator $\omega_0$ as $\hat{\omega}\equiv \omega_0/\gamma$. $\hat{\omega}$ can also be written in terms of the black hole temperature \eqref{Temperature} as $\hat{\omega}=4\pi  T_\text{H}$. This relation differs by a factor of $2$  from the one pertaining to 4D black holes, $\hat{\omega}=2\pi  T_\text{H}$ \cite{Hod:1998vk, Maggiore:2007nq, Schutz:1985km, Nollert:1993zz,  Motl:2002hd, Hod:2005ha, Keshet:2007nv, Panotopoulos:2020mii}. This fact seems to be a peculiarity of black-hole solutions in $2$ and $3$ space-time dimensions \cite{Bhattacharjee:2020nul, Cardoso:2001hn, Birmingham:2001pj, Rincon:2018sgd}. 

Finally, we notice that, in the large-$n$ regime, the QNMs are probing the JT black hole at small distances. This should provide us with information about its microscopic structure, similarly to what happens for the Schwarzschild case. On the other hand, the small-$n$ behavior corresponds to probing large distances, of order of the horizon size. This regime is instead not universal, but depends on $\gamma$, i.e. on the mass of the scalar field and on the form of the coupling function $h(\phi)$ (see the parameter $\alpha$ in the spectra \eqref{freqmassless} and \eqref{freqdilaton}).  

We will use this information in the next two sections, first to give a corpuscular description of the black hole, then to improve such a description by considering the black hole as a statistical ensemble of harmonic oscillators.

\section{Corpuscular description} 
\label{Sec:cd}
A black hole can be considered as a macroscopic quantum system, built up by a large number $N$ of microscopic DOFs \cite{Dvali:2011aa, Dvali:2013eja, Cadoni:2020mgb, Casadio:2015lis, Casadio:2016zpl, Casadio:2021cbv}. The theoretical evidence supporting the validity of this corpuscular description is mounting. Specifically, a black hole can be seen as a coherent quantum  state, which {\it classicalizes} for $N\gg 1$, and satisfies a {\it maximally packaging} condition, relating the maximal amount of DOFs one can pack in the system to its size \cite{Dvali:2011aa, Dvali:2013eja}.  This condition has been shown to be equivalent to the holographic scaling of $N$  \cite{Cadoni:2020mgb}. The possible existence of macroscopic quantum gravity states emerged also in investigations concerning galactic dynamics \cite{Cadoni:2017evg, Tuveri:2019zor, Cadoni:2018dnd}. The latter perspective is further supported by the formulation of a {\it generalized thermal equivalence principle} (GTEP),  which relates the temperature of the statistical ensemble of the $N$ DOFs to the surface gravity of a black hole or to the cosmological acceleration \cite{Tuveri:2019zor}.

Recently, this corpuscular hypothesis gained support also from an apparent relation between the form of the QNMs spectrum of the Schwarzschild black hole and its microscopic description in terms of an ensemble of quantum harmonic oscillators \cite{Cadoni:2021jer}.

This corpuscular interpretation, finds strong support in the Maggiore proposal \cite{Maggiore:2007nq}, according to which in the large overtone number-regime, the proper frequency of the harmonic oscillator describing the QNMs spectrum of the Schwarzschild black hole $\omega = \sqrt{\omega^2_\text{R}+\omega^2_\text{I}}$ can be recast as the dispersion relation of a relativistic particle, i.e. $\omega_n = \sqrt{m^2 + p^2_n}$. Specifically, the particle mass  $m$ is defined in terms  of  the real part of the frequency, $m = \ln 3 \ T_\text{H}$, with $T_\text{H}$ the Hawking temperature of the hole, while the particle momentum $p_n$ is defined in terms of $\omega_\text{I}$ (hence its dependence on the overtone number) and reads $p_n = 2 \pi T_\text{H} \left(n+\frac{1}{2} \right)$. The latter expression, in particular, is pertinent  to a particle quantized on a circle, with length $L = 1/T_\text{H}$, with antiperiodic boundary conditions.
Applying the same reasoning to the QNMs spectrum of the JT black hole \eqref{freqQNMs} considered in this paper, we can rewrite it in the form of a dispersion relation for a relativistic particle with vanishing rest mass 
\beq 
    \omega_n  = p_n,  \quad p_n=  \frac{4\pi}{\beta_\text{H}}\left(n+\gamma \right),
\end{equation}
%.
where now $\beta_\text{H}=1/T_\text{H}$ is the  inverse of the black hole temperature \eqref{Temperature}. As in Ref. \cite{Maggiore:2007nq}, the form of $p_n$ is that pertaining to a particle described by a wave function, $\Psi$, quantized on a circle of length 
$\beta_\text{H}/2$, with boundary conditions 
\beq
\Psi\left(x+\frac{\beta_\text{H}}{2}\right)= e^{{2\pi \text{i}} \gamma}\Psi(x).
\end{equation}
In a corpuscular description, we can therefore consider the JT black hole as a coherent state representing a classical configuration localized in a region of size $r_\text{H}$, with occupation numbers $n_j(p)$ sharply peaked around the characteristic QNMs frequency $p=\hat{\omega}$. Being $\hat{\omega}\propto T_\text{H}$, this is fully in agreement with a thermal, coarse grained, description of the system, according to which the statistical occupation number distribution should be peaked at energies  around $T_\text{H}$.

 Since the system is expected to be weakly coupled in the regime $N\gg 1$, on a first approximation we can neglect the interactions.  We can consequently consider the JT black hole as a bound state of free particles, each with typical energy $\hat{\omega}= 4 \pi T_\text{H}$. This yields
\beq\lb{corp_mass}
M=N \hat\omega,
\end{equation}        
where $M$ is the black hole mass.  

\Cref{Temperature,Entropy} give the scaling of $M$ and $S$ with the temperature $T_\text{H}$:  $M= 2\pi^2 L T_\text{H}^2,\,\, S= 4\pi^2 L T_\text{H}$, typical of a two-dimensional CFT. This is, therefore, a manifestation of  the AdS/CFT correspondence for JT gravity \cite{Cadoni:2000fq, Cadoni:2000kr}.  By using \cref{corp_mass} together with these scaling relations, one can now easily find that the black hole entropy $S$ is proportional to $N$
\beq\lb{corp}
S\propto  N.
\end{equation}       
This is fully consistent with the corpuscular description of the JT black hole as a bound state of $N$ particles, whose thermodynamic entropy roughly scales as the number of DOFs. 

In view of the discussion of the next section, it is quite interesting to compare the behavior of  the JT and of the Schwarzschild black holes. As remarked above, in the latter case, the QNMs spectrum, in the large-$n$ limit, has the same form  given in \cref{freqQNMs},  with $\hat\omega=1/(2 \rH)=  2\pi T_\text{H}$ and $\gamma=1/2$, where now $T_\text{H}=1/(4\pi \rH)$, corresponding to a particle quantized on a circle of length $\beta_\text{H}$, with antiperiodic boundary conditions. The corpuscular interpretation of the Schwarzschild black hole is quite the same as the JT case: it can be regarded as a coherent bound state of $N$ particles with occupation numbers sharply peaked at $\hat\omega\sim T_\text{H}$, such that $M=N \hat\omega$ and $S\propto N$ on a first approximation. 

On the other hand, the scaling of $M$ and $S$ with the Hawking temperature $T_\text{H}$ is quite different: $M=1/(8\pi G  T_\text{H}),\, S=1/(16\pi G T_\text{H}^2)$. This behavior reflects the fact that we do not have a thermal CFT, dual to the black hole, in the Schwarzschild case. This can also be seen as a consequence of the presence of two different length-scales in the two gravity theories. In JT gravity, the gravitational coupling constant is dimensionless and we have an external length-scale, i.e. the AdS length $L$. Conversely, in general relativity, the gravitational coupling constant has dimensions of length squared and  we have no external length-scale. In the next section, we will show that these features  have a deep impact on the black hole description in terms of harmonic oscillators.

Let us conclude this section by noticing an intriguing relationship between our QNMs-motivated corpuscular description of black holes and the recently  proposed GTEP \cite{Tuveri:2019zor}. As already mentioned, the latter explains acceleration in gravitational systems, such as the surface gravity $\kappa$ for black holes and the cosmological acceleration for the de Sitter universe, as a universal macroscopic effect of a large number of thermalized quantum gravity DOFs. The proportionality relation $\hat{\omega}\propto T_\text{H}$ we found for both the Schwarzschild and the JT black holes also implies $\hat\omega \propto \kappa$. This result, therefore, could be seen not as a mere coincidence, but as a consequence of the GTEP.

\section{The JT black hole as statistical ensemble of oscillators}
\label{Sec:BHQuantumOscillator}

In the previous section, we gave a  corpuscular description of the JT black hole in terms of a coherent  state of a large number $N$ of particles, with occupation numbers sharply peaked around the characteristic quasi-normal frequency $\hat\omega$. In this section, we will  give  support  to this microscopic picture of the black hole. Motivated by the form of the QNMs  spectra \eqref{freqmassless} and \eqref{freqdilaton}, we will model the black hole  as a statistical (canonical) ensemble of $N$ harmonic oscillators of frequency  $\hat\omega$. We will follow here the same procedure adopted in Ref. \cite{Cadoni:2021jer} for the Schwarzschild black hole.     

Similarly to the Schwarzschild case, we are mainly interested in macroscopic black holes, i.e. in their large-mass ($M\to \infty$) behavior. There is, however, a crucial difference between the four-dimensional Schwarzschild and the JT black hole. In the latter case, the large-$M$ limit corresponds to large temperatures $T_\text{H}$, whereas in the former it corresponds to the small-temperature regime. This is a simple consequence of the scaling of $M$ in terms of $T_\text{H}$, discussed in the previous section. In the JT case, the large-$M$ behavior will be independent, at leading order, from the vacuum, i.e from the value of $\gamma$ in \cref{freqQNMs}. 

The spectrum will be dominated by the linear behavior in the overtone number $n$, so that  in the large mass limit we can effectively model the JT black hole as an ensemble of $N\gg 1$ decoupled harmonic oscillators. At first sight, modelling the black hole as a set of free particles may seem at odds with  the naive intuition  of a  black hole as a strongly coupled gravitational system. However, one should keep in mind  that the system is strongly coupled in terms  of the 't Hooft coupling  $\lambda= \alpha N$ (see \cref{secint}), but it is weakly coupled in terms of the running coupling constant $\alpha$ \cite{Dvali:2020wqi, Dvali:2011aa, Dvali:2010jz, Dvali:2010bf}.

We adopt a description in terms of a canonical ensemble: the black hole is taken to be in thermal equilibrium with its surroundings at temperature $T=1/\beta$. Owing to its \textit{negative} specific heat and being asymptotically flat, a Schwarzschild black hole cannot be in stable thermodynamical equilibrium with a thermal bath. Consequently, the canonical ensemble is ill defined and requires the system to be confined within artificial external boundaries \cite{Gour:1999ta, Hawking:1982dh}. Conversely, the canonical ensemble is perfectly defined in asymptotically AdS space-times, because its asymptotic timelike boundary naturally acts as a confining box. Notice that, although the AdS$_2$ black hole space-time is not geodesically complete for radial null geodesics, one can impose perfectly-reflecting boundary conditions at the asymptotic timelike boundary of the spacetime. Thus, a large AdS$_2$ black hole is kept at thermal equilibrium with its radiation and is therefore naturally described in terms of the canonical ensemble. \\
In addition, motivated by the fact that the solutions can be fully characterized by a single observable, the mass $M$, we take the number of oscillators $N$ fixed. Following Ref. \cite{Cadoni:2021jer}, we consider $\beta$ and $\hat\omega$  as \textit{independent} variables, so that the temperature $T$ and the black hole mass $M$ can vary independently. This is justified by the fact that the observer at infinity, who only measures the QNMs, has access to the mass of the black hole only, due to the absence of any chemical potential, whereas the horizon radius can fluctuate quantum mechanically; only local measurements would allow to probe these fluctuations \cite{Gour:1999ta}.

In the canonical ensemble, the partition function of a single oscillator, with spectrum given by \cref{freqQNMs}, is
\begin{equation}
    \mathcal{Z}_1 = \sum_{n=0}^{\infty} e^{-\beta \omega_n} = \frac{e^{\frac{2 \beta r_\text{H}}{L^2} \left(1-\gamma\right)}}{e^{\frac{2 \beta r_\text{H}}{L^2}}-1}.
\end{equation}
We consider the JT black hole as a system of $N$, non-interacting, indistinguishable  oscillators. The total partition function 	mau{is} therefore $\mathcal{Z}_N =\mathcal{Z}_1^N$. The average energy and the entropy are
\begin{subequations}
\begin{align}
&\langle E \rangle = -\partial_\beta \ln \mathcal{Z}_N 
= - \frac{2 N \rH}{L^2} (1-\gamma) + 
\frac{2 N \rH}{L^2}\frac{e^{\frac{2 \beta r_\text{H}}{L^2}}}{e^{\frac{2 \beta r_\text{H}}{L^2}}-1};\\
& S = \ln \mathcal{Z}_N +\beta \langle E \rangle=  -N \ln \left(e^{\frac{2 \beta r_\text{H}}{L^2}}-1 \right) + 
\frac{2 N \rH}{L^2} \beta \frac{e^\frac{2 \beta \rH}{L^2}}{e^\frac{2 \beta \rH}{L^2}-1}.
\end{align}
\end{subequations}
In the large-temperature limit, i.e. $\beta \to 0$, we get:
\begin{equation}
\langle E \rangle \simeq \frac{N}{\beta}+ \frac{2 N\rH}{L^2}  \left(\gamma-\frac{1}{2}\right)+\mathcal{O}(\beta).
\label{meanenergyexp}
\end{equation}
\begin{equation}
S \simeq N - N \ln \left(2 \frac{\rH}{L^2} \beta  \right)+\mathcal{O}(\beta^2).
\label{Sexpanded}
\end{equation}
Notice that, consistently with our previous statement, the vacuum contribution (parametrized by $\gamma$) enters only in the subleading terms in the  $\beta\to 0$ expansion. Thus, for the asymptotic observer, which identifies $T=T_\text{H}$, the leading term in the large-mass expansion gives a purely extensive contribution to the energy, satisfying $\langle E \rangle=T S$, which is completely unaffected by the zero-point energies of the oscillators. This behavior has to be compared with that of the Schwarzschild black hole, for which also the vacuum energy provides a non-negligible contribution to $\langle E \rangle$ in the large-mass limit, consistently with the scaling $M\sim  T_\text{H}^{-1}$ \cite{Cadoni:2021jer}.

The leading term in \cref{meanenergyexp} has to be identified with the the black-hole mass $M$ measured by an observer at infinity, who sees the system at thermal equilibrium at the Hawking temperature \eqref{Temperature}. This fixes the number of oscillators to $N=\beta_\text{H} M=\pi \rH / L$. Plugging this result into the leading term of \cref{Sexpanded} gives

\begin{equation}
    S =\pi \frac{ \rH}{L},
\end{equation}
which is half of the Bekenstein-Hawking (BH) entropy (\ref{Entropy}) of the JT black hole.
The next-to-leading order term in the entropy expansion  \eqref{Sexpanded} is a logarithmic correction, in agreement with several results in the literature (see, e.g. Refs. \cite{Mann:1997hm, Akbar:2010nq, Carlip:2000nv, Mukherji:2002de, Medved:2004eh, Domagala:2004jt, Grumiller:2005vy, Fursaev:1994te, Setare:2006ww, Ghosh:1994wb, Kaul:2000kf, Cadoni:2007vf}). These corrections are positive when expressed in terms of the temperature and have a pure thermodynamic origin, as they arise from the high-temperature expansion, and causes an increase in the entropy, as expected (see, e.g. Ref. \cite{Grumiller:2005vy} for a discussion about thermal corrections to the entropy in general two-dimensional dilatonic models). 

The origin of the mismatch of a factor of $2$ between the leading term in the Gibbs entropy for the $N$-oscillator system and the BH entropy \eqref{Entropy} could be traced back to the peculiar topology of the AdS$_2$ space-time. The JT black hole represents just one of the two disconnected wedges of full AdS$_2$. The BH entropy of the hole can be thought as resulting from tracing out the DOFs on the invisible edge of the full AdS$_2$ \cite{Cadoni:2007vf, Maldacena:2001kr}. This means that, in our microscopic description in terms of harmonic oscillators, we should double $N$, keeping however the black hole mass unchanged. In this way, we find $S=2N= 2\pi \rH/L$, matching exactly the BH entropy (\ref{Entropy}).     

\section{Quasi-normal modes and Conformal Symmetry}
\label{Sec:Mapto1DQMmodel}
In the previous sections, we derived and discussed the QNMs spectrum for the JT black hole and used it to build a microscopic description both in terms of a coherent  state of a large number of particles  and as   a statistical ensemble of  harmonic oscillators. Until now, this perspective has been motivated only by the form of the QNMs spectrum \eqref{freqQNMs}. On the other hand, it is well known that  in the AdS/CFT framework, JT gravity allows for a dual description in terms of a CFT. In two spacetime dimensions, the correspondence has both a bulk realization in terms of a 2D CFT \cite{Cadoni:2000fq, Cadoni:2000kr} and an holographic one in terms of conformal quantum mechanics living on the 1D time-like boundary of AdS$_2$ \cite{Strominger:1998yg, Navarro-Salas:1999zer, Cadoni:1999ja, Cadoni:2000ah}. It is therefore tempting to see to what extent the microscopic description of the JT black hole in terms of  harmonic oscillators finds support in the dual 1D boundary, quantum mechanical description. This will be the subject of this section.

AdS$_2$ gravity induces a conformally invariant dynamics on the 1D space-time boundary at spatial infinity $r \to \infty$  \cite{Cadoni:2004mn, Cadoni:2000gm}. Specifically, the boundary theory has the same form of the conformal quantum mechanics proposed by de Alfaro, Fubini and Furlan (DFF) in \cite{deAlfaro:1976vlx}, coupled to an external source. The boundary dynamics is generated by the asymptotic symmetry group of AdS$_2$, the Diff$_1$ group (time reparametrizations), namely the set of transformations which leave the following  asymptotic expressions of the metric and the dilaton invariant
\begin{equation}\label{AsymptSymm}
    g_{tt}\sim -\frac{r^2}{L^2} + \gamma_{tt}(t)+\mathcal{O}\left(\frac{1}{r^2} \right), 
    \,\,\, 
    g_{rr} \sim \ \frac{L^2}{r^2} +\frac{L^4\gamma_{rr}(t)}{\ r^4}+\mathcal{O}\left(\frac{1}{r^6} \right), 
  \,\, \, 
    \phi \sim \left(\rho(t) \frac{r}{L} + \frac{L\gamma_{\phi}(t)}{2 r} \right) + \mathcal{O}\left(\frac{1}{r^3} \right),
\end{equation}
where $\gamma_{tt}$, $\gamma_{rr}$, $\gamma_\phi$ and $\rho$ represent the boundary and the dilaton deformations, respectively. The equations of motion, stemming from the action \eqref{JTnoanomaly}, together with \cref{AsymptSymm}, give the dynamical equations
\begin{equation}\label{boundarydynamicsystem}
    \frac{\ddot{\rho}}{L^2} -\rho \gamma +\beta=0, \qquad \dot{\rho} \gamma + \dot{\beta}=0,    
\end{equation}
where $\gamma \equiv \gamma_{tt}-\gamma_{rr}/2$, $\beta \equiv \rho  \gamma_{rr}/2 + \gamma_\phi$ and the dot indicates differentiation with respect to $t$.

We can now use the Diff$_1$ gauge freedom to fix $\gamma=\text{constant}$ in \cref{boundarydynamicsystem}, and get the equation for the harmonic oscillator 
\begin{equation}\label{rhoddot}
    \ddot{\rho}=\frac{2 \gamma}{L^2} \rho.
\end{equation}
$\gamma$ is a function of the ADM mass \eqref{ADMmass}, $\gamma =-M L$, and depending on its sign, \cref{rhoddot} describes an harmonic oscillator with real frequency ($M<0$), a free particle ($M=0$) or an harmonic oscillator with imaginary frequency ($M>0$), signalizing the presence of dissipative effects. Specifically, in the  black hole solution ($M>0$), \cref{rhoddot} describes an harmonic oscillator with frequency $\omega= \text{i} \hat{\omega}$, with $\hat{\omega}$ given by \cref{freqQNMs}. This result corroborates the microscopic description of the JT black hole in terms of harmonic oscillators, based on the QNMs spectrum, from the dual holographic theory point of view.  

The gauge fixing, leading  to \cref{rhoddot}, breaks the full  Diff$_1$ invariance group, leaving unbroken only the time-translation subgroup. This is a quite strong condition, because one would like to keep unbroken at least the isometry of  AdS$_2$, which is isomorphic to SL(2,R). This group is generated  by the transformation
\begin{equation}\label{Goperator}
    G = u\mathcal{H}+v \mathcal{D}+w\mathcal{K}
\end{equation}
where $\mathcal{H}$, $\mathcal{D}$, and $\mathcal{K}$  generate translations, dilatations and conformal transformations, respectively.  

The spectrum of $G$ can be characterized by the sign of the determinant $\Delta \equiv v^2-4uw$. Specifically, for $\Delta <0$, $G$ is compact, its spectrum is discrete and bounded from below and its eigenstates are normalizable. For $\Delta =0$, the spectrum is continuous and bounded from below. For $\Delta >0$, $G$ is non-compact and its spectrum is unbounded from below. 
The action, invariant under the conformal transformations generated by $G$, takes the form \cite{deAlfaro:1976vlx}  
 \begin{equation}\label{newaction}
     \mathcal{S} = \frac{1}{2}\int d\tau \left(\dot{q}^2+\frac{\Delta}{4}q^2 -\frac{g}{q^2} \right).
 \end{equation}
The  parameter $\Delta$ in the action (\ref{newaction}) plays the same role of the parameter $\gamma$ in \cref{rhoddot}. 

The case of interest in our discussion is $\Delta > 0$, i.e. a non-compact $G$ with an unbounded-from-below spectrum, which corresponds to the $\gamma>0 $ behavior, i.e. the harmonic oscillator with \textit{imaginary} frequencies $\omega= \text{i} \hat{\omega}$, dual to the JT black hole space-time. This is a consequence of the fact that this space-time is endowed with an event horizon and we can describe it in terms of a thermal quantum harmonic oscillator at the horizon temperature \eqref{Temperature} \cite{Cadoni:2004mn}. 

The last point shows, in particular, that the black hole space-time is holographically dual to a 1D conformally symmetric \textit{thermal} system on the boundary, which is of course perfectly consistent with the AdS/CFT correspondence. 

The next step is to look for a correspondence between the spectrum of the DFF boundary operator \eqref{Goperator} on the boundary and the JT QNMs spectrum in the bulk. This is quite natural also in consideration of the fact that QNMs are dual to the response of a thermal system to external perturbations on the boundary \cite{Horowitz:1999jd}. In view of the holographic nature of the correspondence, we expect a matching of the spectra in the $r\to \infty$ ($r_\ast\to 0$) limit  of the bulk radial coordinate. 

Let us then focus on the eigenvalue equation for $G$ \cite{deAlfaro:1976vlx}
\begin{equation}\label{Schrodingerlikeequation}
    \left(-\frac{d^2}{dx^2}+W(x)\right)\psi = 2\mathcal{G}\psi,\quad  W(x) = \frac{g}{x^2}-\frac{\Delta}{4}x^2
\end{equation}
where $2\mathcal{G}$ are the eigenvalues and $\psi$ the eigenfunctions. \Cref{Schrodingerlikeequation} has the form of a time-independent Schr\"odinger equation, with potential  given by $W(x)$.

The spectrum of the eigenvalues and the normalizability of the eigenfunctions are therefore determined by the coupling $g$ in the $x\to 0$ region. Specifically, $g>0$ provides an infinite repulsive well, which keeps the particle confined in the internal region $0<x<\infty$. On the other hand, $g<0$ corresponds to an attractive potential, giving unphysical solutions. The uncoupled case $g=0$ provides non-normalizable eigenfunctions, coherently with the free particle, whose wavefunction is defined on all the line $-\infty < x<\infty$.

The parameter $\Delta$ rules, instead, the behavior at $x \to \infty$. If $\Delta >0$, the potential is monotonically decreasing as $x\to \infty$ and is unbounded from below. If $\Delta =0$, the potential approaches to zero in this limit, thus producing a continuous spectrum bounded from below. If $\Delta <0$ and the coupling $g$ is positive, the potential has a minimum at $x_{\text{min}} = \left(2g/|\Delta|\right)^{1/4}$, and goes to $+\infty$ both for $x\to 0$, $\infty$. It means that the eigenfunctions will be renormalizable and the eigenvalues spectrum discrete and bounded from below. Specifically, this spectrum is equivalent to the harmonic oscillator one with \textit{real} frequencies, with eigenvalues scaling linearly with an integer $n$.

Let us now compare the potential $W(x)$ in \eqref{Schrodingerlikeequation} with the QNMs potential of \eqref{Vrast}. In general, the two potentials  will be different. However, as already anticipated, the relationship with the asymptotic symmetries, discussed at the beginning of this section, implies that the QNMs and the operator $G$ spectra are expected to match in the limit $r\to\infty$ ($r_\ast \to 0$).

The $r_\ast \to 0$ expansion of the potential \eqref{Vrast} gives
\begin{equation}\label{Vinfinity}
    V(r_\ast) =\frac{\a(1-\a)}{3}\frac{\rH^2}{L^4} +\frac{\a(\a+2)}{4r^2_\ast}+ \frac{\a(4\a-7)}{15}\frac{\rH^4 r_\ast^2}  {L^8} + \mathcal{O}(r_\ast^3).
\end{equation}
In order to make a direct comparison with the DFF model, we now define the dimensionless variable
\begin{equation}
    x^2 \equiv  \frac{\rH^2 r_\ast^2}{L^4},
\end{equation}
which brings the asymptotic potential \eqref{Vinfinity} into the form
\begin{equation}
    V(x) = \frac{\rH^2 }{L^4}\left[ \frac{\a(1-\a)}{3} + \frac{\a(\a+2)}{4x^2}+\frac{\a(4\a-7)}{15}x^2\right].
\end{equation}

Comparing now the potential $V(x)$ with the DFF one $W(x)$ of \cref{Schrodingerlikeequation}, we can easily identify the DFF couplings $\Delta$, $g$ in terms of the dilaton coupling parameter $\alpha$: 
\begin{subequations}
\begin{align}
g&=\frac{\a(\a+2)}{4}, \label{secondcorrespo}\\
    -\frac{\Delta}{4}&=\frac{\a(4\a-7)}{15}. \label{thirdcorrespo}
\end{align}
\end{subequations}

Notice that, apart from these identifications, the matching between the QNMs and the DFF operator eigenevalue equations requires also a shift of the values of $\omega^2$  by a $\alpha$-dependent term. Interestingly, however, this shifting term becomes zero when $\alpha =1$, i.e. using the most natural coupling $h(\phi) = \phi$ (see \cref{Sec:QNMSAnalyticalSolutions}).

The sign of $\Delta$ determines whether the frequencies are real or imaginary. We see that, for dilaton couplings $0<\a<7/4$, $\Delta >0$, and we have a correspondence with a thermal harmonic oscillator, with imaginary frequencies on the boundary. If $\a >7/4$, $\Delta <0$ and we have the spectrum of a standard harmonic oscillator, with \textit{real} frequencies.

\Cref{secondcorrespo,thirdcorrespo} clearly show that a non-trivial dilaton coupling $h$ in the Klein-Gordon equation \eqref{KG} is needed to have also a non-trivial boundary dynamics. $h=\text{constant}$ implies $\a \to 0$, so that both $g\to 0$ and $\Delta \to 0$, yielding a free-particle boundary dynamics. Correspondingly, the form of the bulk wavefunction is not compatible with the QNMs boundary conditions. On other hand, we see that, as long as $\a >0$, $g$ is positive, so no unphysical eigenstates are present.

\subsection{Eigenvalue problem for the DFF model with $\Delta >0$}

We showed above that, in the asymptotic $r_\ast \to 0$ regime, the QNMs spectrum for the JT black hole can be put in correspondence with the  eingevalue problem for the DFF operator $G$ with $\Delta >0$. Let us, therefore, consider the eigenvalue equation \eqref{Schrodingerlikeequation} in our specific case. As we saw before, in this case  the spectrum of $G$ is unbounded from below. From a quantum mechanical point of view, these states do not have a physical interpretation and were therefore excluded in the analysis of Ref. \cite{deAlfaro:1976vlx}.  

\Cref{Schrodingerlikeequation} can be solved in terms of special functions. The solution reads
\begin{equation}
\begin{split}
\label{solutionboundary}
    \psi(x) =  A \ 2^{b-\frac{1}{4}}\ e^{-\frac{z}{2}}\ \left(\frac{\text{i}z}{\sqrt{\Delta}} \right)^{\frac{b}{2}-\frac{1}{4}}  \ \mathcal{U}\left(a,b,z \right)+B \ 2^{b-\frac{1}{4}}\ e^{-\frac{z}{2}}\ \left(\frac{\text{i}z}{\sqrt{\Delta}} \right)^{\frac{b}{2}-\frac{1}{4}}  \ \mathcal{L}_{-a}^{b-1}\left(z \right)
\end{split}
\end{equation}
where $A$ and $B$ are two integration constants, $\mathcal{U}(a, b, z)$ is the confluent hypergeometric function, $\mathcal{L}^\a_n(z)$ are the generalized Laguerre polynomials, and we have defined
\begin{equation}
    a\equiv \frac{-4i\mathcal{G} \sqrt{\Delta}+2\Delta+\Delta\sqrt{1+4g}}{4\Delta},\qquad b\equiv1+\frac{\sqrt{1+4g}}{2}, \qquad z\equiv-\frac{\text{i}}{2}\sqrt{\Delta}\ x^2.
\end{equation}
Let us now impose the boundary conditions. The correspondence between the DFF conformal quantum mechanics and QNMs in the bulk, discussed in the previous subsection, implies that, to solve the former, we have to use here the same boundary conditions at $x=0$ and $x=\infty$,  used to solve the latter in the bulk. Regularity of the solution and selection of purely ingoing waves at  $x=\infty$ (corresponding to the horizon in the bulk) requires $B=0$. On the other hand, the confluent hypergeometric functions diverge in $x=0$. Near $x=0$, the first term of \cref{solutionboundary} therefore behaves as (we neglect non-relevant numerical factors) 
\begin{equation}\begin{split}
    x^{b-\frac{1}{2}} \ \mathcal{U}(a, b,z) \sim & \ x^{b-\frac{1}{2}} \frac{\Gamma\left(1-b \right)}{\Gamma\left(1+a-b \right)} +c_1 \ x^{b-\frac{1}{2}} x^{2(1-b)} \frac{\left(\frac{1}{2}-\frac{\text{i}}{2} \right)^{2(b-1)} \Delta^{\frac{1}{2}-\frac{b}{2}}\Gamma\left(b-1 \right)}{\Gamma\left(a \right)}
\end{split}\end{equation}
with $c_1$ a numerical factor. Owning to the fact that $g>0$, the first term goes to zero for $x \to 0$. The second term, however, being proportional to $x^{\frac{1}{2}-\frac{\sqrt{1+4g}}{2}}$, diverges as $x\to 0$. To prevent this, we require the Gamma function to diverge, from which we are able to compute the spectrum of $\mathcal{G}$
\begin{equation}\begin{split}
    &\frac{1}{2}+\frac{\sqrt{1+4g}}{4}-\frac{\text{i}\mathcal{G}}{\sqrt{\Delta}}  =-n, \qquad n=0, 1, 2, ...\\
    &\mathcal{G} = -\text{i}\sqrt{\Delta} \left(n+\frac{1}{2}+\frac{\sqrt{1+4g}}{4} \right).
\end{split}\end{equation}
We see that the eigenvalues are purely imaginary and they scale linearly with $n$, as expected. 

Using \cref{secondcorrespo}, the previous equation becomes
\begin{equation}\lb{DFFS}
    \mathcal{G} = -\text{i}\sqrt{\Delta} \left(n+\frac{1}{2} + \frac{\a+1}{4}\right),
\end{equation}
which has the same form of  the analytic results \eqref{freqmassless}, \eqref{freqdilaton} obtained for QNMs in the bulk, with $\hat{\omega}=\sqrt{\Delta}$. Notice that, however, the expressions for the frequency $\omega_0$  are quite different in the two cases. For QNMs, $\omega_0$ in \cref{freqQNMs} is a function of the horizon radius only (or, equivalently, of the black hole mass), whereas $\omega_0$ depends on $\alpha$ only in \cref{DFFS} (see \cref{thirdcorrespo}).  This discrepancy is due to the fact that the correspondence QNMs/DFF quantum mechanics holds only in the $r\to \infty$ asymptotic regime, where we expect the spectrum to be determined by the perturbation, i.e. the parameter $\alpha$ defining the dilatonic coupling function $h$.

\section{Conclusions}
\lb{sect7}
We investigated the QNMs spectrum for external massless and massive scalar perturbations in the gravitational background of  2D JT black holes. Specifically, we extended previous calculations \cite{Bhattacharjee:2020nul, Cordero:2012je} of the quasi-normal frequencies to the case of massive scalar perturbations. We have shown in this way that they are purely imaginary and scale linearly in the overtone number for all kinds of physical perturbations allowed in two dimensions. This is reminiscent of the energy spectrum of a quantum harmonic oscillator. 

Motivated by this and  extending previous results for the Schwarzschild black hole \cite{Cadoni:2021jer}, we proposed a microscopic corpuscular description of the 2D black hole  in terms of a  coherent  state of $N$ massless particles quantized on a circle and with  occupation numbers sharply peaked on the characteristic QNMs frequency $\hat\omega$. This allowed us to recover the scaling relations $M=N\hat\omega \propto T_\text{H}^2$ and $S \propto T_\text{H}$ of the mass and entropy of the black hole. Such a behavior is typical of a two-dimensional CFT, and it is also perfectly consistent with the quasi-normal frequency being regarded as the inverse of the time-scale required by the conformal theory on the boundary to reach thermal equilibrium \cite{Horowitz:1999jd}. Moreover,  the two scaling relations  imply  linear scaling of the entropy in the number of degrees of freedom, $S \propto N$, consistently with  a corpuscular interpretation. Intriguingly, these results can also be considered as a consequence of the generalized thermal equivalence principle (GTEP), first proposed in Ref. \cite{Tuveri:2019zor}, which explains macroscopic effects, such as accelerations,  in gravitational systems (e.g. black holes, de Sitter universe) as a universal effect of a large number of thermalized quantum gravity degrees of freedom.

We further sharpened this description by modelling the black hole as a statistical ensemble of $N$ free harmonic oscillators  of characteristic frequency $\hat\omega$. Motivated both by the fact that an observer at infinity, who only measures the QNMs, has access to the mass of the black hole only, and by the fact that the time-like AdS boundary acts as a confining box for the space-time, we worked in the canonical ensemble, by keeping the number of oscillators fixed. We focused on the thermodynamic limit of high-temperature, corresponding to high-mass black holes.  The leading-order contribution to the Gibbs entropy yields the Bekenstein-Hawking one, while the sub-leading term is a logarithmic contribution. We also found that $N$ equals to the BH entropy.

We further corroborated our microscopic description by establishing an holographic correspondence between QNMs in the two-dimensional bulk and the DFF conformally-invariant quantum mechanical model. We explicitly showed that, near the AdS boundary, the Schr\"odinger-like equation for the QNMs in the bulk reduces to the eigenvalue equation for the non-compact DFF generator $G$. We also compute the spectrum of $G$, reproducing the linear scaling of the frequency with the overtone number $n$, in agreement with our results in the bulk.

The results of this paper, together with those of Ref. \cite{Cadoni:2021jer}, strongly suggest that modelling a black hole both as a coherent  state of particles with an appropriate dispersion relation and as a statistical ensemble of harmonic oscillators holds universally in the large-mass and large-$n$ limit. This is essentially a consequence of the linear asymptotic scaling of the QNMs frequency in terms of both the overtone number and the black-hole temperature.

\bibliography{QNMEntropy15-11}
\bibliographystyle{ieeetr}

\end{document}